\documentclass[prb, superscriptaddress, footinbib, twocolumn,floatfix]{revtex4}
\usepackage[english]{babel}
\usepackage[dvips]{graphicx}
\usepackage{epsfig}
\usepackage{psfrag}
\usepackage{amsmath}
\usepackage{amssymb}
\usepackage{subfigure}
\begin{document}
\title{Parametric resonances in electrostatically interacting carbon nanotube arrays}
\author{A. Isacsson}
\email{andreas.isacsson@chalmers.se}
\author{J. M. Kinaret}
\affiliation{Department of Applied Physics, Chalmers University of
Technology, SE-412 96 G{\"o}teborg, Sweden}
\begin{abstract}
  { We study, numerically and analytically, a model of a
    one-dimensional array of carbon nanotube resonators in a
    two-terminal configuration.  The system is brought into resonance
    upon application of an AC-signal superimposed on a DC-bias
    voltage.  When the tubes in the array are close to each other,
    electrostatic interactions between tubes become important for the
    array dynamics. We show that both transverse and longitudinal
    parametric resonances can be excited in addition to primary
    resonances. The intertube electrostatic
    interactions couple modes in orthogonal directions and affect the
    mode stability.}
\end{abstract}
 
\maketitle
\section{Introduction}
During recent years, several experimental realizations of
nano-electromechanical (NEM) resonators based on carbon nanotubes
(CNT) or carbon nanofibers (CNF) have been
demonstrated\cite{Sazanova,Zettl,Purcell,Herre,Bachtold,Campbell}.
Carbon nanotubes have established themselves as strong material
candidates for use in NEM-resonator systems, partly due to their
favorable mechanical properties\cite{Hierold} such as low mass and
high elastic modulus.  Thus, using CNTs/CNFs allows for operational
frequencies of NEM-resonators that reach into the GHz regime. In the
most recent experiments, the long predicted high quality factors of
the order of $Q\sim 10^3$ have finally been achieved. This makes these
resonators interesting from a technological point of view with
application areas such as tunable RF-filters and fast low power
switching elements\cite{Cleland,Roukes2001,ITRS}.  However, for such
applications, a major drawback is the high impedance levels offered by
single nanotube devices.  The ensuing low power
transduction\cite{Roukes2001} makes integration of such devices with
current state-of-the-art CMOS technology difficult.  One way to
overcome this problem is to construct devices based on parallel
arrays. For arrays it is desirable to know how interactions
betwen elements affect the operation of devices. It is for instance
important to know how closely spaced array members can be placed
without drastic changes in performance.

Apart from the technological incentive to study MEM/NEM arrays, the
problem is also of fundamental interest. The dynamic response of
coupled NEM/MEM-resonator systems in combination with nonlinearities
is known to lead to unexpected and/or unintuitive behavior. Examples
are intrinsic localized
modes\cite{MEMS_Array1,MEMS_Array2,MEMS_Array3,Herre2} and mode
synchronization\cite{Syncro1,Syncro2}. In addition, parametric
resonances in MEM/NEM arrays have been studied both experimentally and
theoretically\cite{Buks1, Buks2,Roukes,LC,Mioduchowski}. In those
studies parametric response was induced by applying an AC-voltage
component between alternating beams in beam-arrays.  Parametric
resonances can be narrower than fundamental resonances and finds use
in for instance parametric amplifiers.

In this paper we study theoretically a vertical one-dimensional
regular array of CNT/CNF resonators (see figure 1). When the
resonators are not too widely separated, electrostatic interactions
between the tubes become important and affect the dynamical response
of the system. While the system considered in this paper shares some
features with previously studied systems\cite{Buks1,
  Buks2,Roukes,LC,Mioduchowski}, there are several important
differences.  Among them two are worthy of special attention. Firstly,
all the tubes in the system are connected to the same voltage source,
containing both a DC- and an AC-component, thereby eliminating the
need for individual contacting of alternating array members. This
makes the interactions between tubes repulsive, rather than
attractive.  Secondly, in contrast to beams with rectangular
cross-sections, where the characteristic vibration frequencies differ
for vibrations in different directions, CNT/CNF have a circular
cross-sections and motion in two dimensions plays an important role.
\begin{figure}[t]
\begin{center}
\epsfig{file=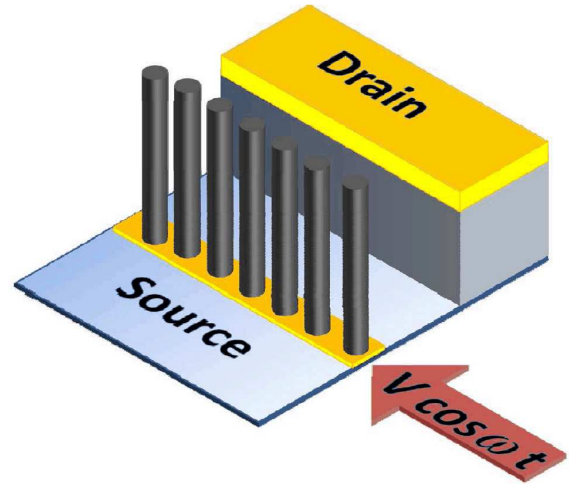,width=8cm}
\end{center}
\caption{
  (Color online) Schematic layout of a vertically oriented carbon
  nanotube NEM array. Through patterning of catalysts, regular one
  dimensional arrays of carbon nanotubes or nanofibers may be grown on
  top of an electrode (Source). A second electrode (Drain) is formed
  from a deposited layer at the same height as the tube tips.
  Actuation and transduction of tube displacement is achieved through
  electrostatic (capacitive) coupling between the tubes and the drain
  electrode.
  \label{fig:system1}}
\end{figure}

We find that the fundamental resonance, where the tubes oscillate in
unison towards the drain electrode, is not drastically affected by
interactions. However, several new resonances appear including
longitudinal resonances, where the tubes oscillate in the direction
along the array. For small arrays, these resonances
have the form of hardening Duffing type resonances which develop into
a band of resonances as the array gets larger. Further, two parametric
resonances are present in the systems. Both transverse as well as
longitudinal motion may be parametrically excited, both with multiple
branches and complex bifurcation structures for the larger arrays. The
electrostatic coupling between tubes also affect the stability of the
longitudinal motion which becomes unstable due to parametric
excitation of transverse oscillations.

We begin this paper by presenting, in section~\ref{sec:model}, a
simplified lumped electromechanical model to derive the main
qualitative features of the system. Then, in
section~\ref{sec:general}, based upon numerical integration of the
equations of motion, the general features of the dynamic response of a
one-dimensional CNT resonator array are discussed. To better
understand the main characteristics of this response we focus on the
the smallest possible array (two tubes) in
section~\ref{sec:two_tubes}.  The two-tube system is treated both
numerically and analytically. We derive frequency response
equations and analyze stability for the various resonances using
perturbation theory.  These analytical results are found to work as
good approximations for finding the loci of the resonances also in
larger systems.  This treatment is then followed up, in
section~\ref{sec:Several}, with a discussion of how the response
changes as the arrays become larger before concluding in
section~\ref{sec:conclusions}.

\section{Model \label{sec:model}}
For an array consisting of $N$ tubes we denote the coordinates of the
central axis of each (undeformed) tube by ${\bf X}_i^0=(X_i,Y_i)$
where $i=1,...,N$.  For the vibrations of the beams we consider only
excitations of the fundamental flexural modes for which a lumped model
is suitable\cite{Cleland, Pelesko}. Describing the position of the tip
of the cantilevers by the coordinates ${\bf X}_i$ we use the equations
of motion\cite{Isacsson_NT}:
$$m_i\ddot{\bf X}_i+m_i\gamma\dot{\bf X}_i+m_i\omega_{0i}^2({\bf
  X}_i-{\bf X}_i^0) ={\bf F}_i^{\rm el.}\quad i=1,..,N.$$
Here $m_i$
are the effective masses of the tubes and $\omega_{0i}$ the natural
resonance frequencies. For tubes with circular cross sections these
values are given by\cite{Isacsson_NT} $m_i=\rho A_i L_i/5.684$ and
$\omega_{0i}=3.516L_i^{-2}\sqrt{EI_i/\rho_i A_i}$ where $A_i$ is the
cross-sectional area, $I_i$ the moment of inertia, $E$ the Young
modulus and $L_i$ the length of tube number $i$.  
We have also introduced a viscoelastic damping term $\gamma_i$ for each
tube to account for mechanical losses. In the absence of a gaseous
medium surronding the tubes, the main sources of dissipation are
clamping losses and Ohmic losses.
 
Aside from elastic forces, external electrostatic forces ${\bf
  F}_i^{\rm el.}$ act on the tubes. These forces depend on the
geometry as well as the instantaneous charge distributions on the
tubes. To find the exact charge distributions on the tubes is a hard
problem.  Numerical simulations using FEM and BEM of electrostatically
interacting tubes\cite{Klas} reveal that the charge is mainly located
at the tip of the tubes.  Furthermore, the main contributions to the
bending moments arise from forces close to the tube tips.  Thus to
model the electrostatic forces we make the simplified assumption that
the charge $Q_i$ on each tube is concentrated to a conducting
spherical shell located at the tip of each tube. The model we consider
is one of conducting spheres, attached with springs to their
equilibrium positions and being able to move in the $xy$-plane.  The
drain electrode, which can be taken to be at zero potential, is
modeled as an infinite conducting plane. There is also a possibility
of actually having metallic grains on top the tubes.  In
plasma-CVD growth of CNT/CNF from Ni catalysts, tip-growth results in
the metallic catalysts residing at the tips of the tubes.

In general, the full charge distribution, rather than the total
charge, on each sphere is needed to find the forces. Provided the
intertube separation as well as the tube-drain separations are larger
than the tube diameters, the dominant contribution to the
electrostatic forces comes from the monopole contributions of these
charge distributions.  Restricting attention to
this case the electrostatic free energy of the system [the tubes being
biased by the common time-dependent voltage $V(t)$] is
$${\cal F}_{\rm el.}={\cal F}_{\rm self}+{\cal F}_{\rm int.}
-V(t)\sum_{i=1}^N Q_i,$$
where the electrostatic self-energy is
$${\cal F}_{\rm self}=\frac{1}{4\pi\epsilon_0}\sum_{i=1}^N \frac{Q_i^2}{D_i}$$
and the total electrostatic interaction energy is
$${\cal F}_{\rm int.}=\frac{1}{4\pi\epsilon_0}\sum_{i=1}^N\sum_{j>i}^N
\frac{Q_iQ_j}{|{\bf X}_i-{\bf X}_j|}
-\frac{1}{8\pi\epsilon_0}\sum_{i,j=1}^N\frac{Q_iQ_j}{|{\bf
    X}_i^\prime-{\bf X}_j|}.$$
Here ${\bf X}_i^\prime=(-X_i,Y_i)$
denote the images of ${\bf X}_i=(X_i,Y_i)$ in the plane $X=0$.  The
charge distribution in the array is then found from solving the linear
system ${\partial {\cal F}_{\rm el.}}/{\partial Q_i}=0$ after which
the electrostatic forces may be found as ${\bf F}_i=-\nabla_{{\bf
    x}_i} {\cal F}_{\rm el.}$.

We consider now a uniform system with
identical tubes, i.e.,  $D_i=D$,
$\omega_{0i}=\omega_0$, $\gamma_i=\gamma$ and $m_i=m$. Rescaling 
the coordinates to dimensionless form according
to $X_i=Dx_i$, $Y_i=Dy_i$ yields
\begin{equation}
\ddot{\bf x}_i+\gamma\dot{\bf x}_i+\omega_{0}^2({\bf x}_i
-{\bf x}_i^0)=\frac{1}{mD}{\bf F}_i^{\rm el.}.
\label{eq:dyneq}
\end{equation}
In the same way we rescale electric quantities through introducing a
unit voltage $V_0$, \emph{i.e.}  $V=vV_0$, $Q_i=q_i4\pi\epsilon_0 D
V_0$ and the electrostatic charging energy ${\cal E}_{\rm
  C}=4\pi\epsilon_0 D V_0^2$.  The corresponding dimensionless
electrostatic free energy $f_{\rm el.}\equiv {\cal F}_{\rm el.}/{\cal
  E}_{\rm C}$ is then
$${{ f}_{\rm el.}}=\sum_{i=1}^N{q_i(q_i-v)} +\sum_{i=1}^N\sum_{j>i}^N
\frac{q_iq_j}{|{\bf x}_i-{\bf x}_j|}
-\frac{1}{2}\sum_{i,j=1}^N\frac{q_iq_j}{|{\bf x}_i^\prime-{\bf
    x}_j|}.$$

In order to study the response of the system we solve the dynamic
equations~(\ref{eq:dyneq}) numerically using a velocity-Verlet
algorithm. In the next section we present the main qualitative
feutures of the mechanical response of the system to a harmonic
driving field. A more quantitative discussion is then carried out in
sections~\ref{sec:two_tubes} and~\ref{sec:Several}.

\begin{figure}[t]
\begin{center}
\epsfig{file=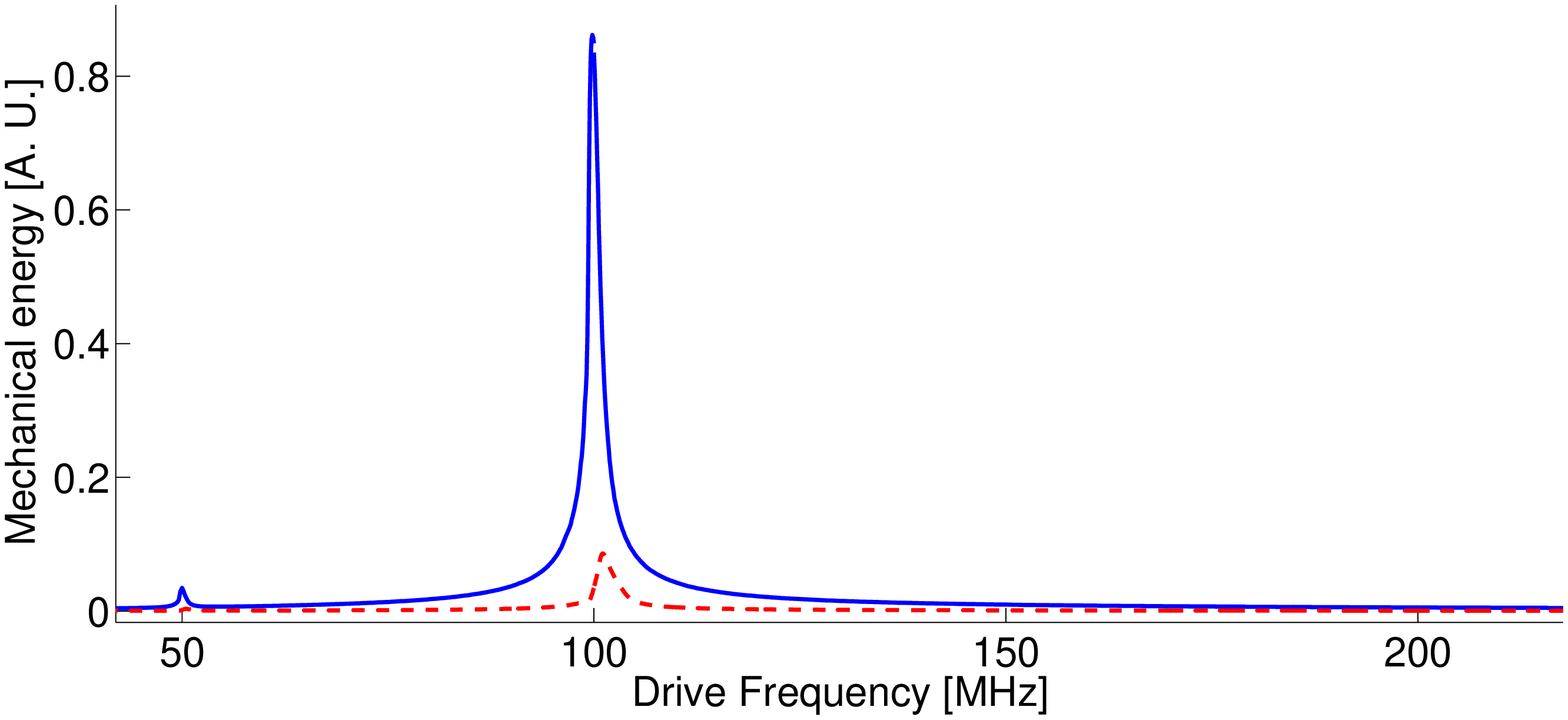,width=9cm}
\epsfig{file=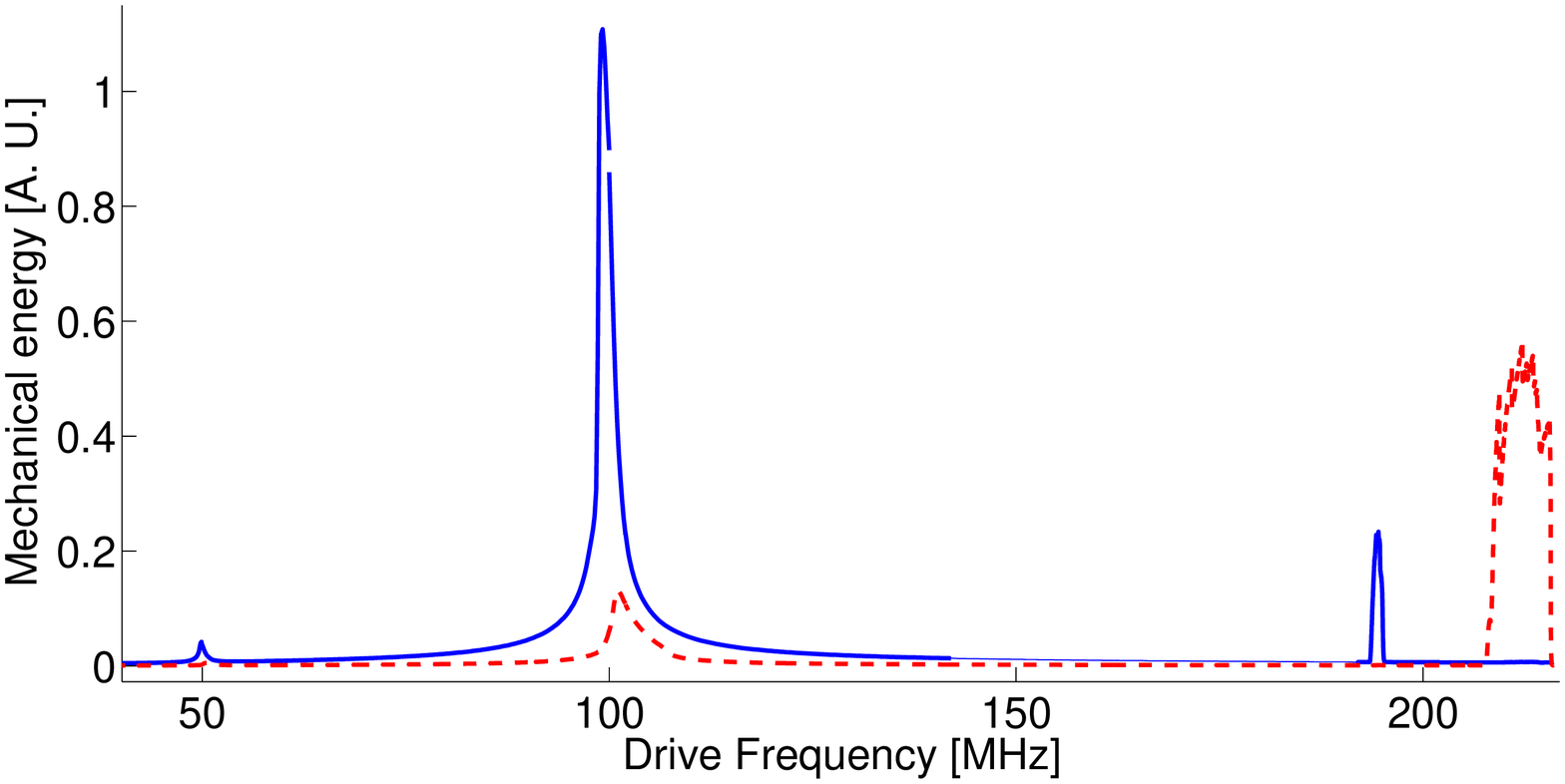,width=9cm}
\epsfig{file=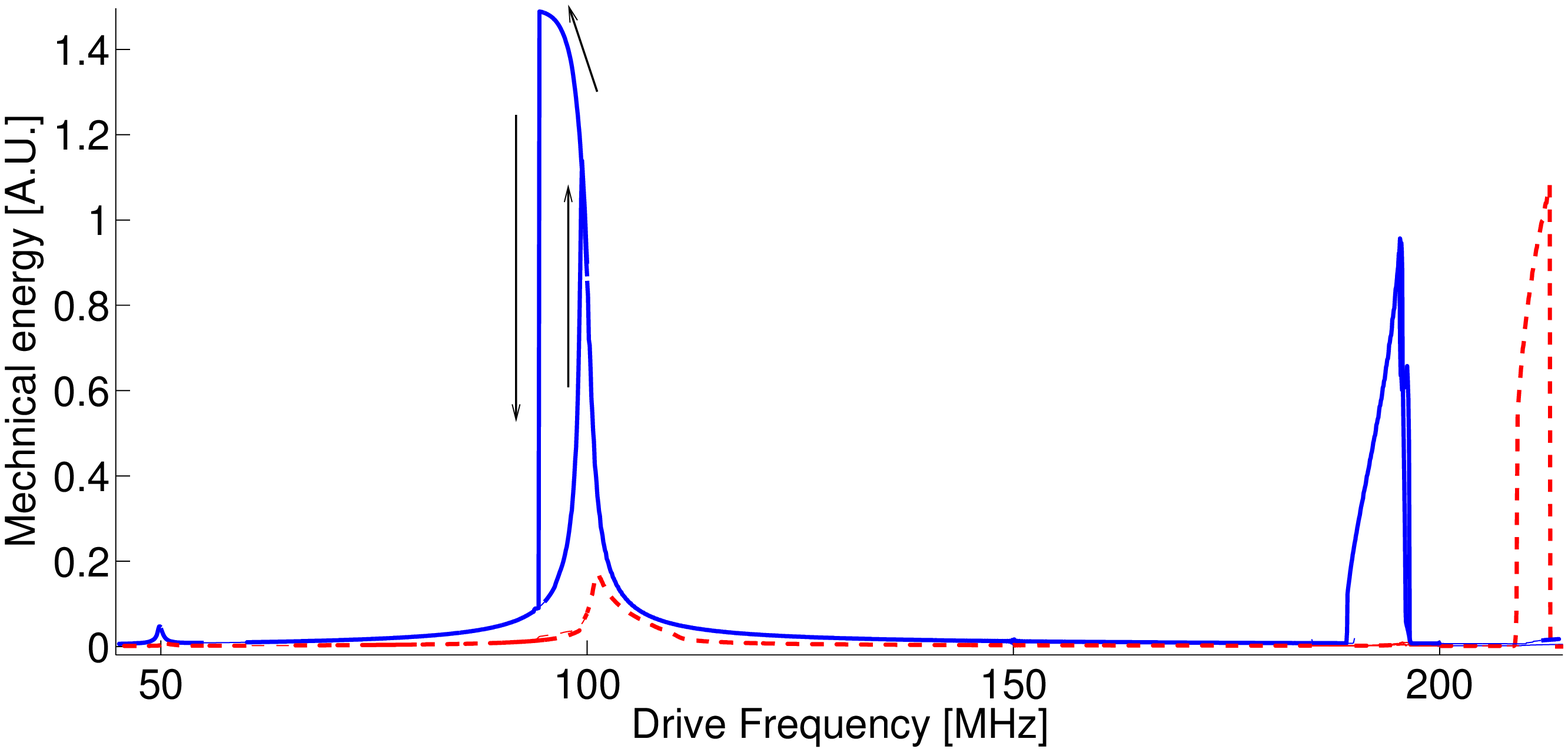,width=9cm}
\end{center}
\caption{
  (Color online) Time averaged mechanical energy for a harmonically driven array
  with 100 tubes. The blue solid lines represent the mechanical energy stored
  in transverse modes (vibrations towards the drain electrode). The red,
  dashed lines represent the energy stored in longitudinal modes (vibration along the
  array axis). Arrows indicate the directions of transitions in the hysteretic region. {\bf
    Top}: Intertube separation 200 nm. {\bf Middle}: Intertube
  separation 150 nm.  {\bf bottom}: Intertube separation 125 nm.
\label{fig:100a}}
\end{figure}
\section{Response to harmonic driving \label{sec:general}}
Figure~\ref{fig:100a} shows the response of an array with 100
tubes when it is driven with an AC-signal on the source in combination
with a static DC-bias voltage $V(t)=V_0+V_1\cos(\omega_D t)$ (DC-bias voltage $V_0=12$ V, AC-signal $V_1=2$ V).
The tubes are each 1 $\mu$m long with a diameter of 25 nm. The
distance to the drain electrode is 150 nm.  Plotted are the two
orthogonal components of the mechanical energy corresponding to
transverse (blue solid line) and longitudinal vibrations (red dashed
lines). The energy is scaled in terms of the dimensionless units
introduced above with a timescale set to $t_0=0.1$ ns.

In the top panel the spacing between the tubes is 200 nm. Clearly
visible is the primary transverse resonance (blue lines) where all the
tubes oscillate in phase with each other.  This mode corresponds to
the resonance of a single nanotube. A band of longitudinal modes can
be seen (dashed line) just above 100 MHz. In the middle panel the
tubes are more closely spaced (150 nm) and two additional resonances
are present.  The transverse (around 190 MHz) is a parametrically
excited resonance where each tube oscillate with half the driving
frequency. In this resonance neighboring tubes oscillate with opposing
phases (optical mode). Above 210 MHz, is another parametric resonance
in the more closely spaced arrays.  This is a parametric
resonance of the band of longitudinal modes of the array. In the
bottom panel the spacing has been narrowed down further to 125 nm. The
parametric resonances are now stronger and the primary transverse
resonance has become hysteretic. The appearance of hysteresis can here be
understood by considering the attractive force between a tube and the
drain electrode.  For widely separated tubes, each tube is attracted
by its own image potential alone, while for more closely spaced tubes, the
images charges from neighboring tubes contribute to this force.

In the next section we show how these resonances and their main
characteristics can be understood from analyzing a two-oscillator
array.  Then, in section~\ref{sec:Several} we study, numerically, how
the response changes qualitatively as the size of the arrays grow
larger.  The parametric resonances, both the transverse and the
longitudinal, show a complex behavior with multiple bifurcation
points. These are not shown in the panels of figure~\ref{fig:100a} but
will be addressed further in section~\ref{sec:Several}.

\section{Two oscillators, Case study \label{sec:two_tubes}}
In this section we study the simplest case, namely an array consisting
of only two cantilevers with identical physical parameters. This case
can be analyzed analytically and serves to validate numerical modeling
and provides insights for larger arrays.

We take the positions of the undeflected tubes to be ${\bf
  x}_1^0=(x_0,y_0)$, ${\bf x}_2^0=(x_0,-y_0)$ where $x_0<0$ and
$y_0>0$ and the drain to be the plane $x=0$. Introducing the variables
$x_{\pm}=x_1\pm x_2$ and $y=y_1-y_2$ we have the equations of motion
\begin{eqnarray}
\ddot{x}_++\gamma \dot{x}_++\omega_0^2 (x_+-2x_0)=\nu\omega_0^2 v^2\left[\frac{g_1^2}{(x_++x_-)^2}\right.\nonumber&&\\
+\left.\frac{g_2^2}{(x_+-x_-)^2}-2g_1g_2\frac{x_+}{(x_+^2+y^2)^{3/2}}\right]&&\label{eq:twotub1}\\
\ddot{x}_-+\gamma \dot{x}_-+\omega_0^2 x_-=\nu\omega_0^2v^2\left[\frac{g_1^2}{(x_++x_-)^2}\right.&&\nonumber\\
-\left.\frac{g_2^2}{(x_+-x_-)^2}+2g_1g_2\frac{x_-}{(x_-^2+y^2)^{3/2}}\right]\label{eq:twotub2}&&\\
\ddot{y}+\gamma \dot{y}+\omega_0^2 (y+2y_0)=2\nu\omega_0^2v^2 g_1g_2\left[\frac{y}{(x_-^2+y^2)^{3/2}}\nonumber\right.&&\\
-\left.\frac{y}{(x_+^2+y^2)^{3/2}}\right].&&\label{eq:twotub3}
\end{eqnarray} 
The relevant electromechanical coupling constant is $\nu={{\cal E}_{\rm C}}/{{\cal E}_{\rm mech.}}=
{{\cal E}_{\rm C}}/{mD^2\omega_0^2}$. In terms of numbers $\nu\approx10^{-3}V_0^2L^3D^{-5}$ if length is measured in
nm and the tubes are assumed solid with a Young modulus of 1
TPa\cite{oneTPa}. The functions $g_{1,2}$ are found by solving
exactly the electrostatic problem and are given by
\begin{eqnarray}
g_{1}&=&(x_++x_-)\frac{(x_+-x_-)(\Delta-2)-1}{(x_+^2-x_-^2)(\Delta^2-4)-4x_+-1},\nonumber\\
g_{2}&=&(x_+-x_-)\frac{(x_++x_-)(\Delta-2)-1}{(x_+^2-x_-^2)(\Delta^2-4)-4x_+-1}\nonumber
\end{eqnarray}
where $\Delta$ is defined as
$$\Delta=\frac{1}{\sqrt{x_-^2+y^2}}-\frac{1}{\sqrt{x_+^2+y^2}}=\Delta_--\Delta_+.$$

\begin{figure}[t]
\begin{center}
\epsfig{file=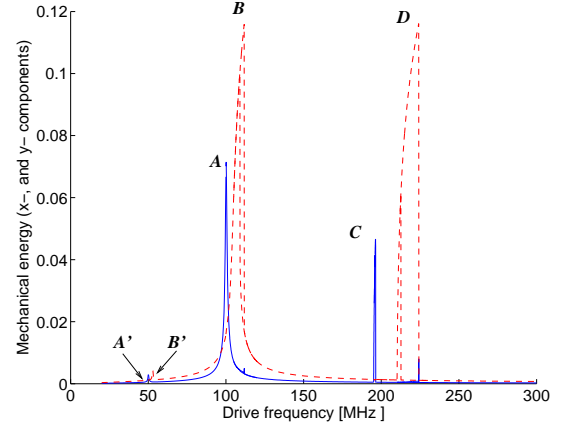,width=8cm}
\end{center}
\caption{
  (Color online) Average mechanical energy of a two-tube system in
  response to harmonic driving. {\bf Blue solid lines:} Average
  mechanical energy stored in vibrations in the $x$-direction.  {\bf
    Red dashed lines:} Average mechanical energy stored in vibrations
  in the $y$-direction. Both upwards and downwards sweeps in frequency
  were made.  {\bf A, A'}: Fundamental resonances in the
  $x$-direction. Both tubes oscillate in phase towards the drain
  electrode ($x_+$-resonance).  {\bf B, B'}: Fundamental resonances in
  the $y$-direction. The tubes oscillate with opposing phases in the direction
  parallel to the drain electrode ($y$-resonance).  {\bf C}:
  Parametric resonance in the $x$-direction. The tubes oscillate with opposite
  phases at half the driving frequency in the
  direction towards drain electrode ($x_-$-resonance).  {\bf D}:
  Parametric resonances in the $y$-direction. The tubes oscillate with opposing
  phases at half the driving frequency in the
  direction parallel to the drain electrode ($y$-resonance).
\label{fig:sim_2_tubes}}
\end{figure}

A spectrum that reveals the most important features of the response to
a harmonic AC-drive on the gate is shown in
figure~\ref{fig:sim_2_tubes}. This figure was obtained from numerical
integration of the dynamic equations for a system with the following
parameters: tube diameter $D=25$~nm; tube lengths $L=1\mu$m; bare
quality factor $\kappa_0=100$; tube positions
$(X_0,Y_0)=(-150,62.5)$~nm; Young modulus $E=1$~TPa; tube density;
$\rho=1.2$~g/cm$^3$. The applied voltage to the system was $V_0=12$~V
and $V_1=2$~V.  

The figure was obtained by sweeeping the drive frequency both upwards
and downwards. On the vertical axis of figure~\ref{fig:sim_2_tubes}
the dimensionless average mechanical energy of the tubes is shown.
The motions in the longitudinal direction ($y$-direction) and the
transverse direction ($x$-direction) have been separated for clarity.
Both the transverse response (blue line) as well as the longitudinal
response (red) show three main peaks each.  We have labeled these
peaks $A,A',C$ and $B, B', D$ respectively.  Hysteresis in the
frequency plane is present in the peaks $B$, $C$ and $D$ (for the peak
$C$ the hysteresis is too narrow to be clearly seen in
figure~\ref{fig:sim_2_tubes}). In the subsections below we treat each
of these resonances in more detail. Note that the subsection labels
follow the labelling of the peaks in figure~\ref{fig:sim_2_tubes}.

We begin by determining the stationary points.
For small deflections around equilibrium it is sufficient to keep
only the dominant terms in $1/x_+$ and $1/y$ which yield the new
dynamic equations:
\begin{eqnarray}
\ddot{x}_++\gamma \dot{x}_++\omega_0^2 (x_+-2x_0)=\nu\frac{\omega_0^2 v^2}{2}\left[\frac{1}{x_+^2}-\frac{x_+}{(x_+^2+y^2)^{3/2}}\right]&&\nonumber\\
\ddot{y}+\gamma \dot{y}+\omega_0^2 (y+2y_0)=\nu\frac{\omega_0^2v^2}{2}\left[-\frac{1}{y^2}-\frac{y}{(x_+^2+y^2)^{3/2}}\right]&&\nonumber
\end{eqnarray}
\begin{eqnarray}
\label{eq:stab_pt}
\end{eqnarray} 
In the limit of large intertube separation ($y\rightarrow \infty$) one
retains the result of noninteracting tubes whereas the
limit $x\rightarrow\infty$ reduces the problem to one in the
$y$-direction only. The equations also decouple in the limit of small
$|x_+|$ (recall that both $x_+$ and $y$ are negative) due to screening
of the electrostatic interaction between the tubes by the drain
electrode.
The system~(\ref{eq:stab_pt}) can be used to determine the stationary
deflections $x_+(t)=x_s$, $y(t)=y_s$ in the absence of an
AC-component.  These time-independent solutions are found by solving
the system
\begin{eqnarray}
x_s&=&2x_0+\frac{\nu}{2}\left[\frac{1}{x_s^2}-\frac{x_s}{(x_s^2+y_s^2)^{3/2}}\right]\nonumber\\
y_s&=&-2y_0+\frac{\nu}{2}\left[-\frac{1}{y_s^2}-\frac{y_s}{(x_s^2+y_s^2)^{3/2}}\right].
\end{eqnarray}
For small biases $v_0$ (i.e.
$\nu\ll1$) we solve pertubatively in $\nu$ and get
\begin{eqnarray}
x_s&\approx&2x_0\left[1-\frac{\nu}{16y_0^3}\left(\sigma_0^{-3}+\Sigma_0\right) \right]\nonumber\\
y_s&\approx& -2y_0\left[1+\frac{\nu}{16y_0^3}\left(1-\Sigma_0\right)\right],
\label{eq:xsys}
\end{eqnarray}
where $\sigma_0=|x_0/y_0|$ and $\Sigma_0\equiv(1+\sigma_0^2)^{-3/2}$.
The term $\Sigma_0$ accounts for mutual screening of the tubes (if $\sigma_0=|x_0/y_0|\rightarrow 0$ then $\Sigma_0\rightarrow
1$). This approximation is valid for small static deflections and far
away from snap-in (the tubes making contact with the drain).  Note
that the system~(\ref{eq:xsys}) rests on the approximation $|1/x_+|\ll
1$ and $|1/y|\ll 1$. This is consistent with the approximation of only
keeping the monopole contribution to the total charge distribution on
the tube tips.

\subsection{Fundamental transverse resonances, ($x_+$-resonance)}
In the fundamental transverse resonances  ($A$ and $A'$ in
figure~\ref{fig:sim_2_tubes}) the tubes oscillate in phase
with each other. Neither of these resonances differ
appreciably in nature from those of single tube systems.  The
subharmonic arises from the double frequency component of the driving
occuring due to the $v^2$ term. The main resonance tunes downwards in
frequency with increasing bias and has a Duffing type nonlinearity of
the softening kind. We omit the analysis of this resonance in this
paper since it has been already thoroughly studied previously in the
litterature in conjunction with single cantilever resonators (see for
instance Ref.~\onlinecite{Isacsson_NT}).

\begin{figure}[t]
\begin{center}
\epsfig{file=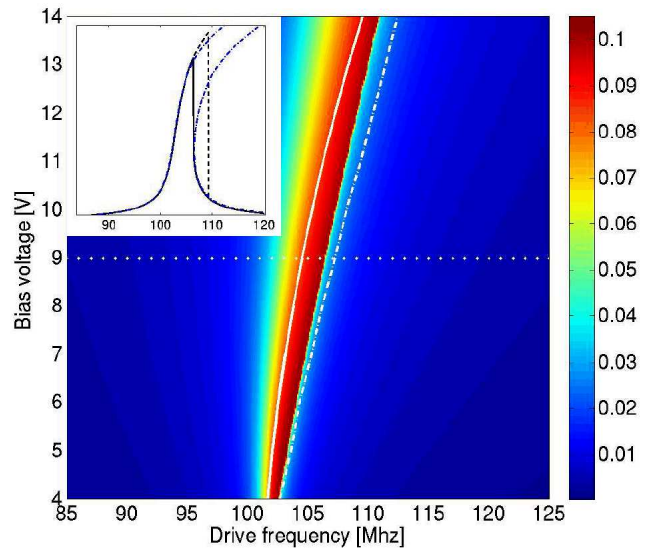,width=9cm,clip}
\end{center}
\caption{ 
  (Color online) False color plot of mechanical energy of the primary
  longitudinal resonance ($y$-resonance) as a function of bias voltage
  (vertical axis) and drive frequency (horizontal axis). The figure
  was created sweeping the frequency downwards. The {\bf
    solid white line} corresponds to the expression~(\ref{eq:wcy}) for
  the central frequency $\omega_c^{(y)}$. Sweeping frequency downwards
  results in an abrupt change in the response at the bifurcation
  frequency $\omega_B^{(y)}$ (the high-frequency edge of the large
  amplitude (red) region). The {\bf dashed white line} shows the the
  bifurcation point found from solving the frequency response
  equation~(\ref{eq:freqresp1}). The {\bf inset} shows the response
  along the cross section at a bias of 9 V (along the dotted line). The
  thick black curves come from numerical simulations
  whereas the dash-dotted blue lines are solutions to the
  frequency response equation~(\ref{eq:freqresp1}).
  \label{fig:y_res}}
\end{figure}

\begin{figure}[t]
\begin{center}
  \epsfig{file=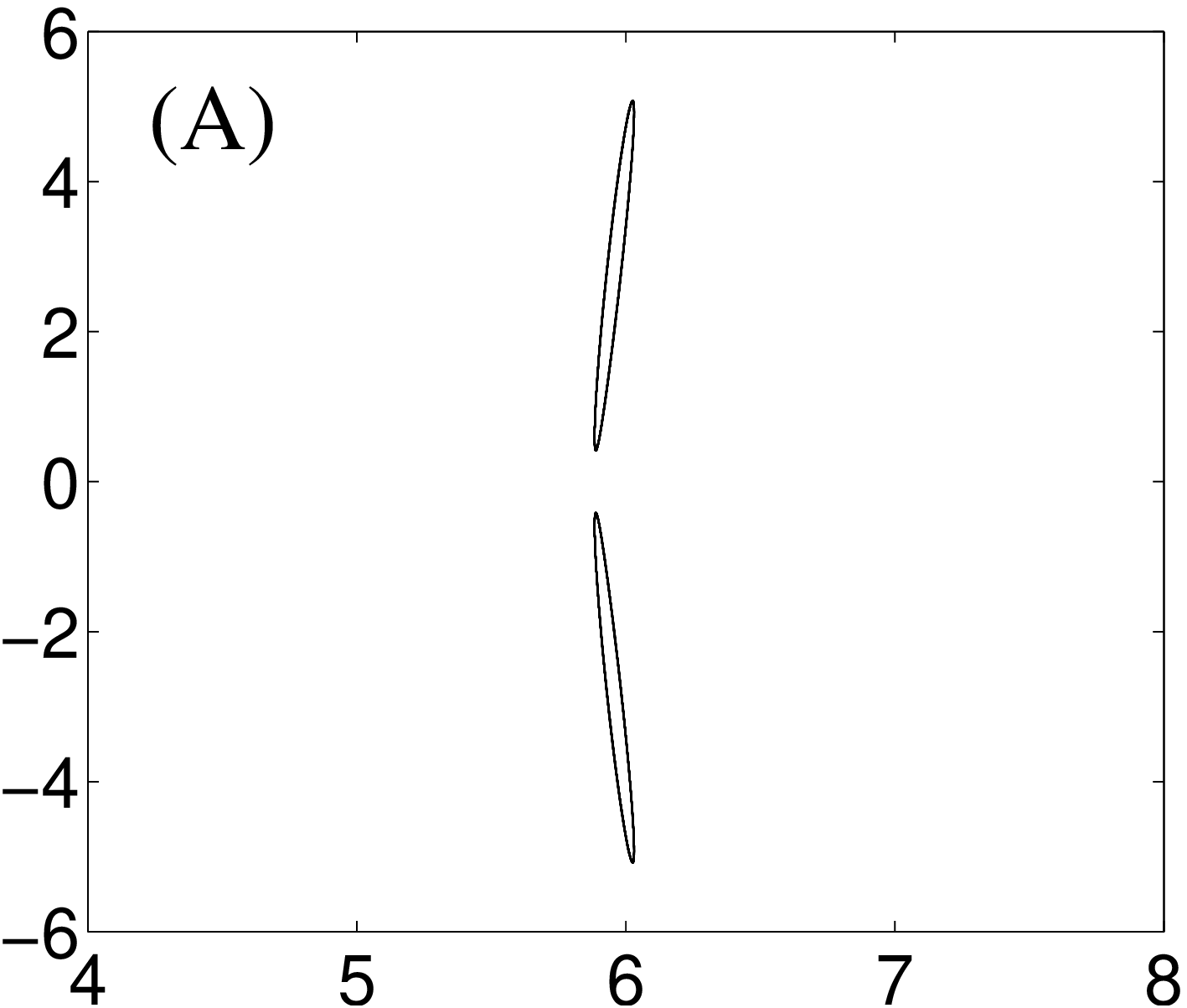,width=4cm,clip}\epsfig{file=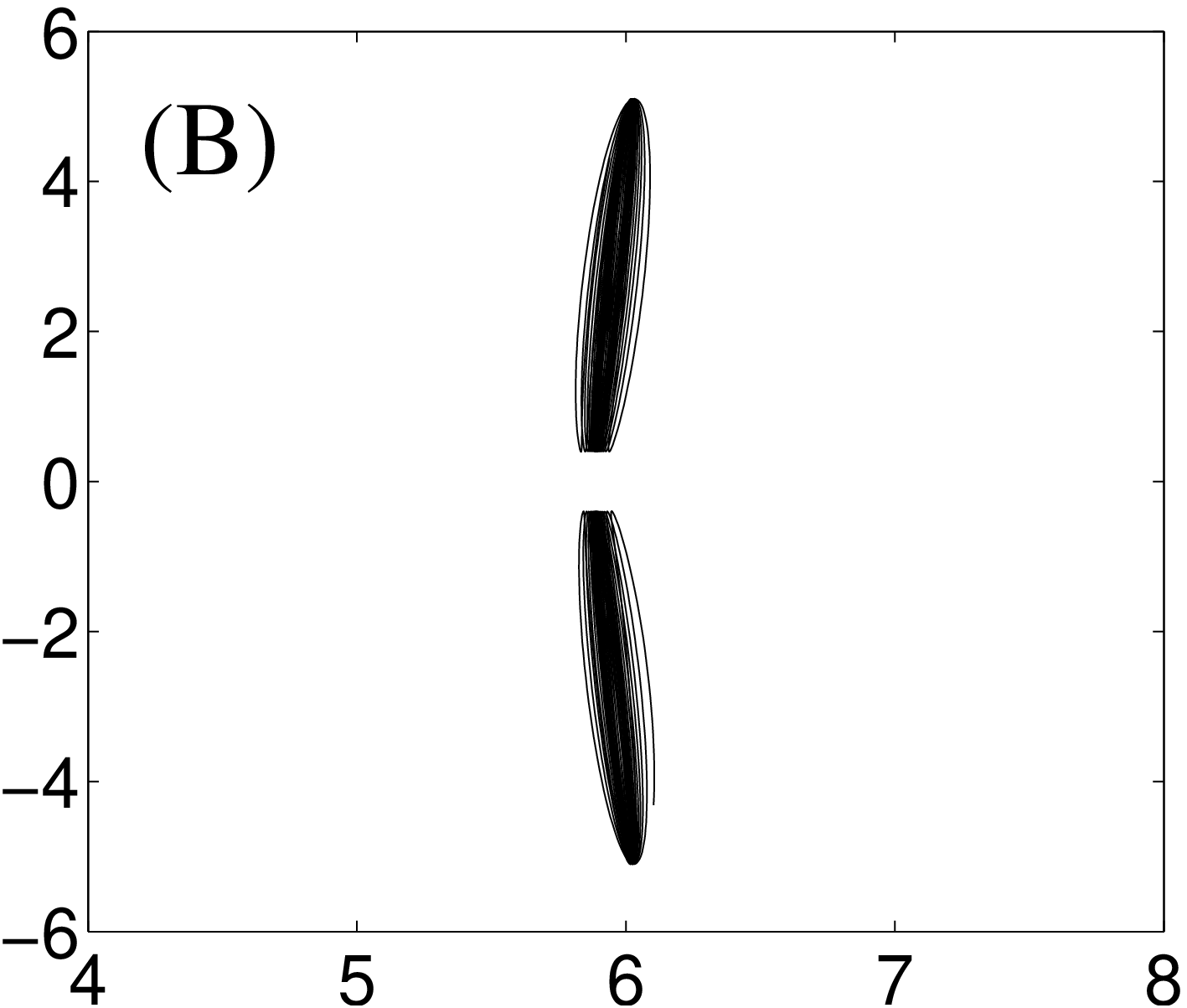,width=4cm,clip}\\
  \epsfig{file=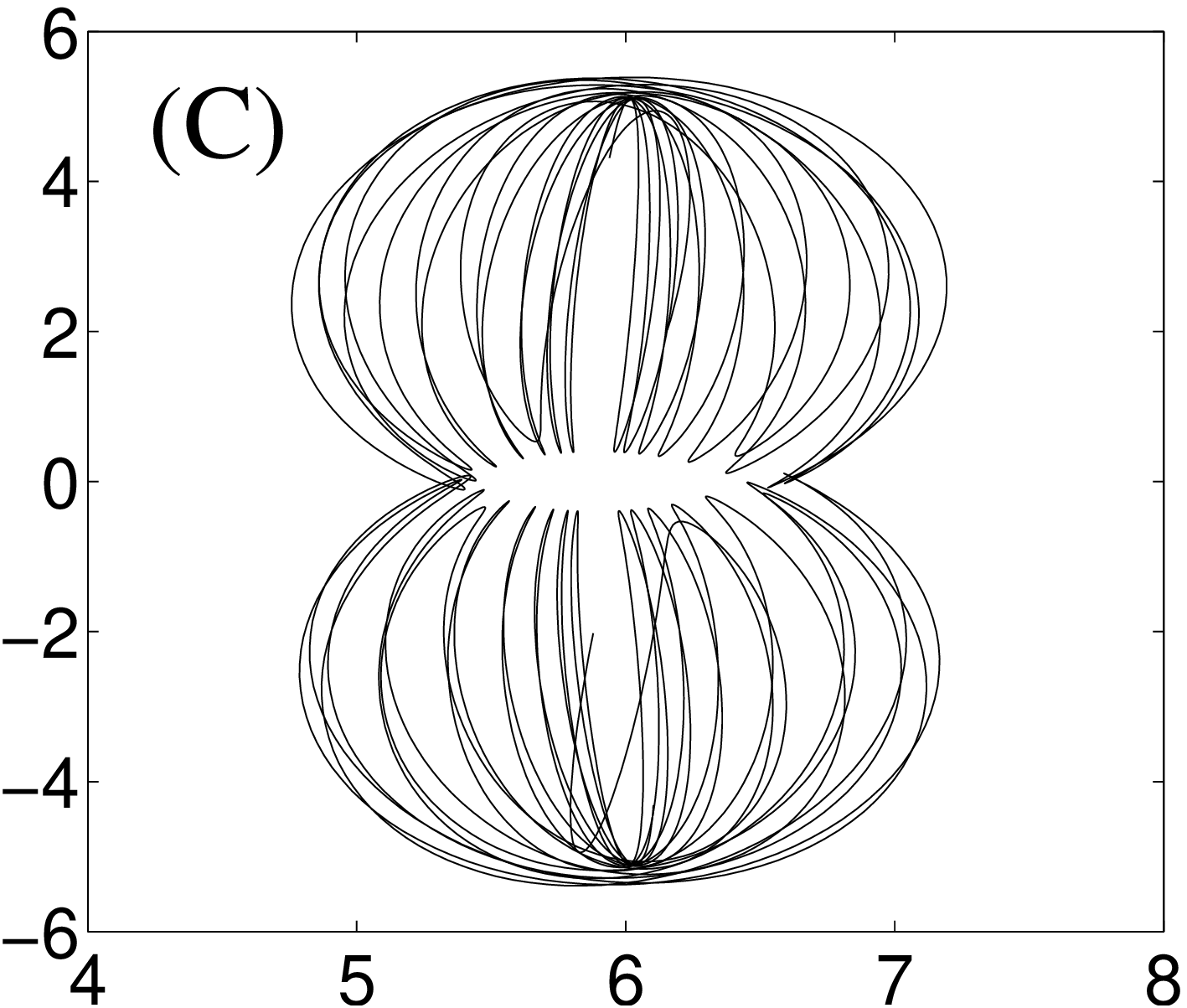,width=4cm,clip}\epsfig{file=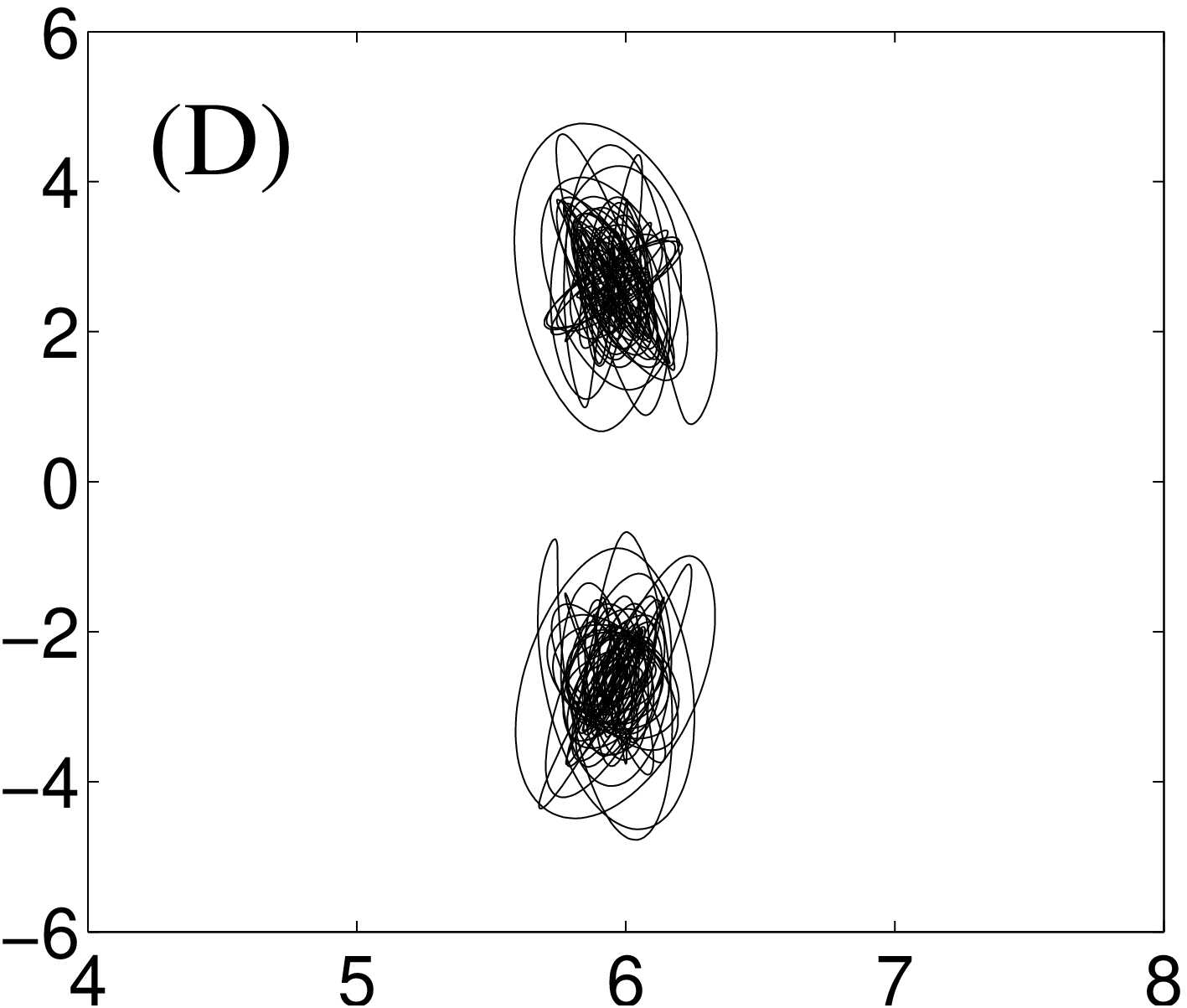,width=4cm,clip}\\
  \epsfig{file=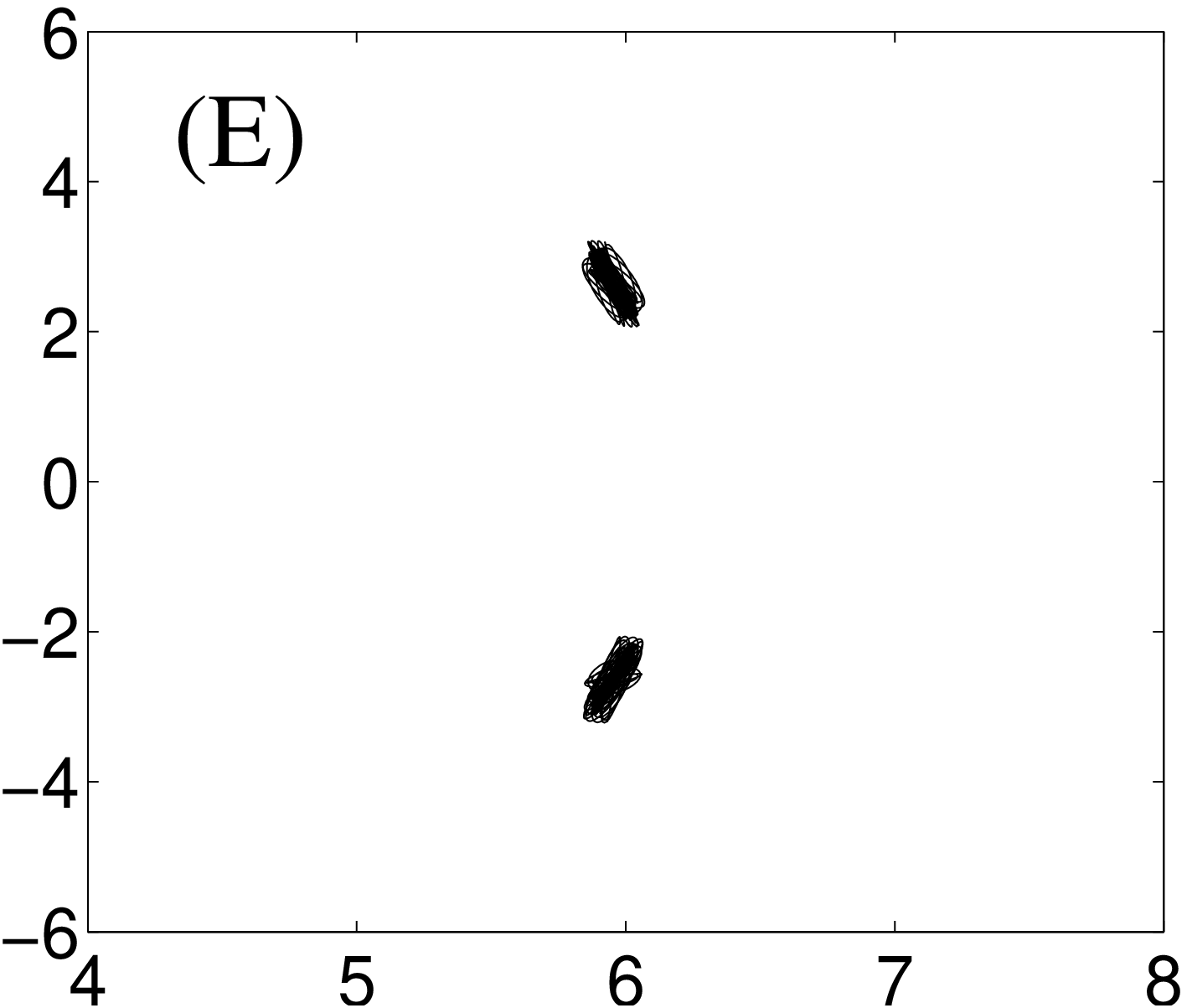,width=4cm,clip}\epsfig{file=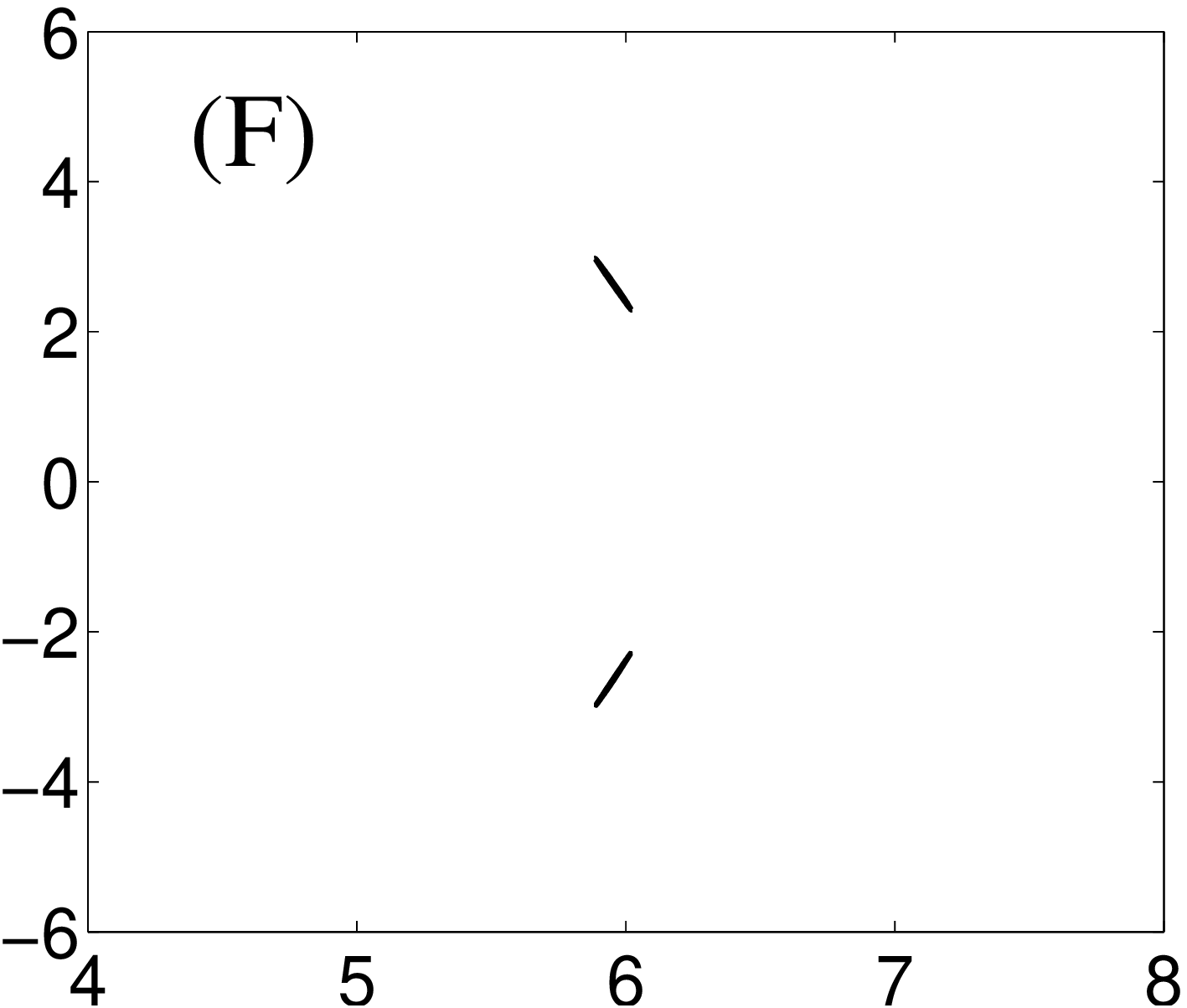,width=4cm,clip}\\
\end{center}
\caption{
  Mechanism for destabilization of the high amplitude branch of the
  fundamental $y$-resonance ({\bf B}-resonance in
  figure~\ref{fig:sim_2_tubes}).  The figures show the time evolution
  of the trajectories of the tubes in the $xy$-plane during
  destabilization.  {\bf A:} Initially the tubes are executing high
  amplitude motion in the $y$-direction. The coupling to the motion in
  the $x-$ direction causes instability of the $x_-$ mode which starts
  to grow ({\bf B}). After having reached high amplitude in both $x-$
  and $y-$ directions ({\bf C}) irregular motion ensues ({\bf D})
  before the motion finally settles in the low amplitude branch of the
  $y$-resonance ({\bf F}).
  \label{fig:ydestab}}
\end{figure}

\subsection{Fundamental longitudinal resonance, ($y$-resonance)}
The fundamental $y$-resonance (resonance $B$ in
figure~\ref{fig:sim_2_tubes}) has the shape of a Duffing resonance
with a hardening nonlinearity.  A more clear view of the resonance is
seen in the inset of figure~\ref{fig:y_res}. In this figure a false
color plot of the mechanical energy in the primary longitudinal
resonance ($y$-resonance) as a function of bias voltage (vertical
axis) and drive frequency (horizontal axis) is shown. The figure was
created sweeping the frequency downwards. Clearly visible is the
upwards tuning of resonance frequency with increasing bias and the
sharp onset of resonance at the bifurcation point.  The inset shows
the response along the 9V bias cross-section (along the
dotted line). The thick black curves are the results from numerical
simulations of the dynamic equations~(\ref{eq:dyneq}). The solid black
line correspond to downard frequency sweep while the dashed line to
upward frequency sweep.

The characteristics of the fundamental $y$-resonance can be found
using perturbation theory.  Considering this 
resonance we take $x_+=x_s$ and assume $x_-=0$.  When $x_-=0$ the
product $g_1g_2$ simplifies to
\begin{equation}
g_1g_2=\frac{1}{[\Delta+2+x_+^{-1}]^2}.
\label{eq:yappro1}
\end{equation}
For small oscillations we may further set
\begin{equation}
1/\sqrt{x_s^2+y^2}\approx 1/\sqrt{x_s^2+y_s^2}=\alpha.
\label{eq:yappro2}
\end{equation}
To obtain an estimate of the parametric dependence of the resonance character we
analyze the system using the method of averaging~\cite{Nayfeh2} by making the Ansatz 
$y(t)=y_s+Y(t)\cos(\omega_D t)$
in response to a drive given by
$v(t)^2\approx v_0^2[1+2\epsilon\cos(\omega_D t+\delta)]$. 
Assuming $\left|\dot{Y}/Y\right|\ll \omega_0$, the differential equation
for the amplitude $Y$ is
\begin{eqnarray}
y_s&=&-2y_0-\frac{\nu}{2}K_0(Y,y_s)\nonumber\\
\dot{Y}&=&\omega_0\nu\frac{\epsilon}{2}\sin\delta\left[K_2(Y,y_s)-K_0(Y,y_s)\right]-\frac{1}{2}\gamma Y\nonumber\\
Y\omega_D&=&Y\omega_0+\frac{1}{2}\omega_0\nu\left(K_1+\epsilon \cos\delta\left[K_0+K_2\right]\right).\label{eq:Yavg}
\end{eqnarray}
Here 
\begin{eqnarray}
K_n(Y,y_s)&\equiv&\frac{2}{\pi}\int_0^{2\pi}\,d\phi \cos n\phi\frac{A^2\left(1+\alpha^3y^3\right)}{(y_s+Y\cos\phi-A)^2}\nonumber\\
&&\label{eq:Kndef}
\end{eqnarray}
with $A\equiv(2+x_s^{-1}-\alpha)^{-1}$.  

For small oscillation
amplitudes where the response does not bifurcate we solve the system
to first order in $\nu$ in the limit $Y\rightarrow 0$. For the
amplitude $Y$ and the center of resonance $\omega_c^{(y)}$ we get
\begin{eqnarray}
Y&=&\frac{\omega_0\nu{\epsilon}\left(1-\Sigma_0\right)}{8 y_0^2\sqrt{\left(\omega_D-\omega_c^{(y)}\right)^2+\frac{\gamma^2}{4}}}\nonumber\\
\omega_c^{(y)}/\omega_0&=&1+\frac{\nu}{16y_0^3}\left[1+\left(\frac{1}{2}\sigma_0^2-1\right)\Sigma_0^{5/3}\right].\label{eq:wcy}
\end{eqnarray}
Equation~(\ref{eq:wcy}) is useful for estimating the center frequency
even for large oscillations as can be seen in figure~\ref{fig:y_res},
where $\omega_c^{(y)}$ obtained from equation~(\ref{eq:wcy}) is drawn as
the solid white line.  As the drive gets stronger bifurcation occurs
(see figure~\ref{fig:sim_2_tubes}) and the shape of the resonance is
found from solving the frequency response equation
\begin{equation}
\omega_0^2\nu^2\epsilon^2=\frac{\gamma^2 Y^2}{[K_2-K_0]^2}
+\frac{[2Y(\omega_D-\omega_0)-\omega_0\nu K_1]^2}{[K_0+K_2]^2}.
\label{eq:freqresp1}
\end{equation}
The blue dash-dotted lines of figure~\ref{fig:y_res} depict the
resonances obtained from equation~(\ref{eq:freqresp1}), showing good
agreement for small amplitudes.  The locus of the bifurcation point
$\omega_B^{(y)}$ may also be determined from solving
equation~(\ref{eq:freqresp1}) and this solution for $\omega_B^{(y)}$ is
shown as the white dashed line in figure~\ref{fig:y_res}.

\begin{figure}[t]
\begin{center}
\epsfig{file=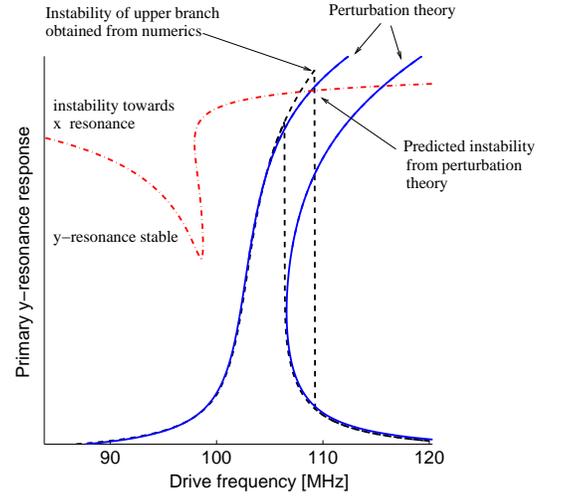,width=8cm,clip}
\caption{
  (Color online) Fundamental $y$-resonance along the dashed line of
  figure~\ref{fig:y_res}. Blue solid lines were obtained from solving
  equation~(\ref{eq:freqresp1}) while the black dashed lines are from
  numerical simulations. Above the red dash-dotted line, perturbation
  theory predicts the fundamental longitudinal resonance to be unstable towards
  parametric excitation of the $x_-$-resonance. The point of
  destabilization of the upper branch occuring where the dash-dotted
  red line and the upper blue line cross.\label{fig:y_destab_pert}}
\end{center}
\end{figure}

Whereas both the location of the resonance and the bifurcation point
can be estimated using Eq.~(\ref{eq:Yavg}), this is not true for
finding the extent of the hysteresis. The destabilization of the high
amplitude branch is connected with an instability towards resonance of
the $x_-$-mode. The process of destabilization is depicted in
figure~\ref{fig:ydestab} where the entire time evolution of the
trajectories of the tubes are shown.  Starting in panel A, the system
is in the high amplitude branch. The instability towards resonance of
the $x_-$-mode causes an increase in motion in the transverse
direction (panels B and C).  Decay to the lower branch (panel F)
occurs through irregular motion of the tubes (panels D and E). The
location where the high-amplitude branch of the $y$-resonance becomes
unstable can be found analytically using perturbation theory.  In
figure~\ref{fig:y_destab_pert} the region of instability towards
parametric excitation of the $x_-$-resonance is shown as the red
dash-dotted line. As can be seen, a good estimate of the locus of the
destabilization can be determined. The perturbative analysis is found
in Appendix~\ref{app:destaby}.


\subsection{Instability towards parametric resonance ($x_-$-resonance)}
We now turn the attention to the parametric resonance of the
$x_-$-mode (the $C$-resonance of figure~\ref{fig:sim_2_tubes}).
Writing out explicitly the right hand side of the equation of motion
(\ref{eq:twotub2}) we have
\begin{equation}
\ddot{x}_-+\gamma\dot{x}_-+\omega_0^2x_-
=-2x_-\omega_0^2 [1+2\epsilon\cos(\omega_D t)]F(x_-)\label{eq:eomxm}
\end{equation}
with 
$$F=\nu\left(\frac{2x_s(\Delta-2)^2-2(\Delta-2)}{[(x_s^2-x_-^2)(\Delta^2-4)-4x_s-1]^2}
-\frac{g_1g_2}{(x_-^2+y_s^2)^{3/2}}\right).$$
The right hand side is proportional to $x_-$ characteristic for a
parametric drive.  A simple parametrically driven harmonic oscillator
$$\ddot{x}+\gamma\dot{x}+\omega_0^2x=\omega_0^2 K[1+2\epsilon\cos(\omega_D t)]x$$
will be unstable\cite{Nayfeh} if $\omega_D^-<\omega_D<\omega_D^+$ where
\begin{equation}
\omega_D^{\pm}=\omega_0\left[2\sqrt{1-K}\pm\sqrt{\frac{K^2}{1-K}\epsilon^2-\kappa_0^{-2}} \right],\quad \kappa_0=\omega_0/\gamma,
\label{eq:parreg}
\end{equation}
provided the discriminant is positive, i.e.
\mbox{$\epsilon>\sqrt{{1-K}}/(\kappa_0 K)$}.
\begin{figure}[t]
\begin{center}
\epsfig{file=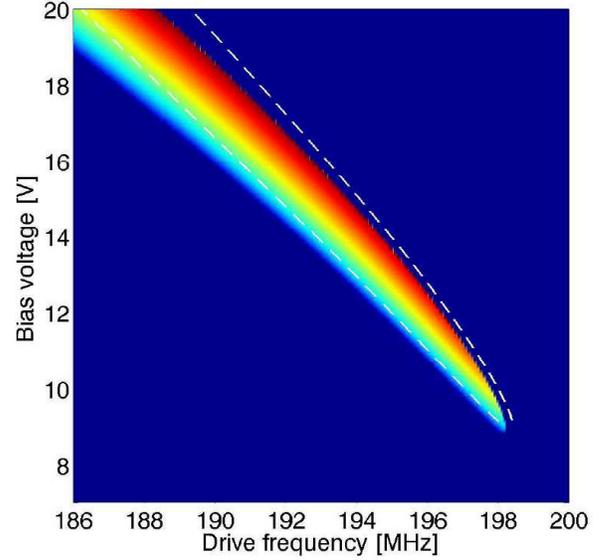,width=8cm}
\end{center}
\caption{
  (Color online) False color plot of the mechanical energy in the
  parametric $x_-$-resonance (resonance $C$ in
  figure~\ref{fig:sim_2_tubes}) as a function of bias voltage
  (vertical axis) and drive frequency (horizontal axis).  The dashed
  white line denotes the region of instability according to equations
  (\ref{eq:parreg}) and (\ref{eq:Kx}).
\label{fig:twotubcomp}}
\end{figure}
To find the point of instability we keep only the lowest order term in
$x_-$ in equation~(\ref{eq:eomxm}). This gives
\begin{equation}
K=2\nu\frac{(2-\Delta_0)(2+\sigma_s^3)x_s+\sigma_s^3}{x_s^2(\Delta_0+2+x_s^{-1})^2},
\label{eq:Kx}
\end{equation}
where $x_s$ and $y_s$ are the stationary points and $\sigma_s\equiv
x_s/y_s$ and $\Delta_0=-\Delta_+-y^{-1}$. A comparison between
numerical simulations and the region of instability is shown in
figure~\ref{fig:twotubcomp}. For small biases the agreement between
theory and numerics is good while it deviates for larger biases. This
deviation is due to the approximate relations (\ref{eq:xsys}) to find
$x_s$ and $y_s$.
\begin{figure}[t]
\begin{center}
\epsfig{file=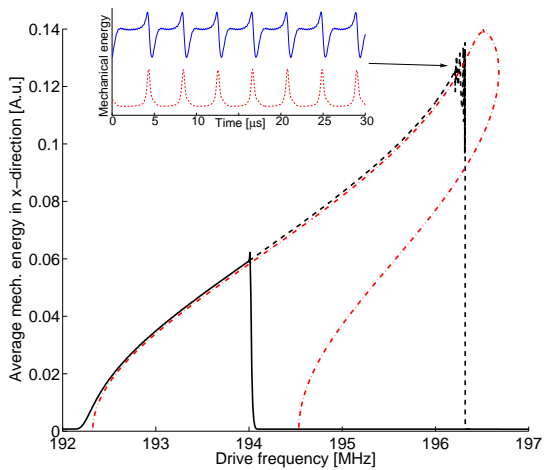,width=8cm}
\end{center}
\caption{
  (Color online) Close-up of the response of the parametric
  $x_-$-resonance (Cross-section at $V_0=14$ V of
  figure~\ref{fig:twotubcomp}).  The resonance has the characteristics
  of a parametrically driven Duffing resonator with a hardening
  nonlinearity. Near the point of instability of the upper branch,
  coupling to the longitudinal mode causes beats 
  where energy is transferred between transverse and longitudinal
  modes periodically in time. Black curves are from numerical
  simulations while the red dash-dotted curves come from solving the
  frequency response equation~(\ref{eq:frespxm}).  The {\bf inset} shows the
  average mechanical energy stored in transverse and longitudinal modes
  as function of time. The red curve, showing energy
  for the longitudinal mode, has been magnified 500 times and
  vertically displaced for clarity.
\label{fig:xparamclose}}
\end{figure}

For larger amplitudes we must consider the full equation of motion
(\ref{eq:eomxm}).  Introducing action angle coordinates
$x_-=X(t)\cos[\phi(t)]$ and $\dot{x}_-=-X(t)\omega\sin[\phi(t)]$, and
expanding $F$ in a Fourier series
$$F(x_-)=F(X\cos\phi)=a_0/2+\sum_{n=1}^\infty a_n\cos n\phi,$$
we obtain after averaging out fast variables the autonomous system
\begin{eqnarray}
\dot{X}&=&X\left[\epsilon\frac{\omega_0^2(a_4-a_0)}{2\omega}\sin\delta-\frac{1}{2}\gamma\right]\nonumber\\
\dot{\phi}&=&\omega+\epsilon\frac{\omega_0^2(a_0+2a_2+a_4)}{2\omega}\cos\delta\nonumber\\
\omega&=&\omega_0^2\sqrt{1+a_0+a_2}.\nonumber
\end{eqnarray}
Here $\delta$ is the relative phase of oscillation with respect to the
drive.  We note that in the limit $X\rightarrow 0$ we have $a_2,a_4=0$
and $a_0=K$.  A comparison between the results of perturbation theory
and numerical simulation is shown in figure~\ref{fig:xparamclose}.
Here the mechanical energy in the parametric $x_-$-resonance is shown
for a bias of $V_0=14 V$ (black solid line is the downward frequency
sweep and dashed line the upward frequency sweep). The red dash dotted
line is the result of solving the frequency response equation
\begin{equation}
\epsilon^2\frac{\omega_0^2}{\omega^2}=\frac{\gamma^2}{(a_4-a_0)^2}+\frac{(\omega_D-2\omega)^2}{(a_0+2a_2+a_4)^2}.\label{eq:frespxm}
\end{equation} 

While agreement between perturbation theory and numerics is good it
does not work well close to the point of instability of the upper
branch. Here, there is noise in the curve obtained from numerical
simulations. This noise comes from coupling to the longitudinal mode.
The inset in figure~\ref{fig:xparamclose} shows how the average energy
stored in transverse (blue) and longitudinal modes (red) vary in time.
The curves have been displaced for clarity and the red curve is
magnified 500 times.  While the energy transferred to the longitudinal
mode is very small compared to the energy in the transverse mode, the
excited longitudinal vibrations has great impact on the transverse vibrations.

\begin{figure}[t]
\begin{center}
\epsfig{file=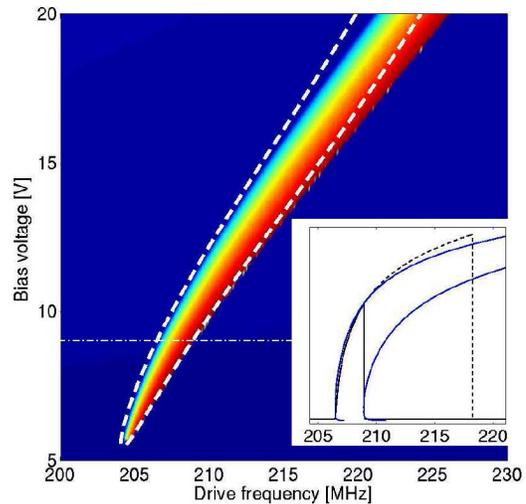,width=7cm,clip}
\end{center}
\caption{ 
  (Color online) False color plot of the mechanical energy in the
  parametric longitudinal resonance ($y$-resonance) as a function of
  bias voltage (vertical axis) and drive frequency (horizontal axis).
  The dashed white line denotes the region of instability according to
  equations (\ref{eq:parreg}) and (\ref{eq:Ky}). The {\bf inset} shows
  the response along the cross section at 9 V bias (dotted line). The
  thick black curves come from numerical simulations and the solid
  blue lines are solutions to the frequency response
  equation~(\ref{eq:freqresp2}).
  \label{fig:y_res_param}}
\end{figure}

\subsection{Parametric longitudinal resonance ($y$-resonance)}
Finally we study the conditions for observing the parametric
longitudinal resonance ($D$-resonance in
figure~\ref{fig:sim_2_tubes}).  As in the case of the parametric
resonance in the transverse direction, the region of instability in
the frequency plane towards parametric resonance in the $y$-direction
is determined by the equation~(\ref{eq:parreg}).  Starting from the
equation~(\ref{eq:twotub3}) and making again the approximations in
(\ref{eq:yappro1}) and (\ref{eq:yappro2}) we find
\begin{eqnarray}
K&\approx& -2\nu A^2\left[\alpha^3-\frac{2}{(y_s-A)^3}\right]\approx\frac{\nu}{y_s^3}\left[1+\frac{1}{2}\Sigma_s\right].
\label{eq:Ky}
\end{eqnarray}
A comparison between numerical simulations and the region of
instability is shown in figure~\ref{fig:y_res_param}. The figure was
created sweeping the frequency downwards and the bifurcation edge is
visible as the sharp transition between dark (blue) and bright (red).
For small biases the agreement between theory and numerics is good
while it deviates for larger biases. This deviation is again due to
using the approximate relations (\ref{eq:xsys}) to find $x_s$ and
$y_s$ respectively.

As in the preceeding subsections we may use perturbation theory to
study the large amplitude response of the parametric resonance.
Assuming $y(t)=y_s+Y\cos(\omega_D t/2)$ and $v(t)^2\approx
v_0^2(1+2\epsilon\cos[\omega_D t+\delta])$ the frequency response
equation can be derived
\begin{eqnarray}
&&\frac{Y^2\left(1+\frac{\nu K_1}{Y}-\frac{\omega_D^2}{4\omega_0^2}\right)^2}{(K_1+K_3)^2}=\nu^2\epsilon^2-\frac{\omega_D^2Y^2}{4\omega_0^2\kappa_0^{-2}(K_3-K_1)^2}\nonumber\\
&&\label{eq:freqresp2}\\
&&\tan\delta=\frac{\gamma Y\omega_D(K_1+K_3)}{(K_3-K_1)([\omega_0^2-\omega_D^2/4]Y-\omega_0^2\nu K_1)}
\end{eqnarray}
where $K_n$ are given by equation~(\ref{eq:Kndef}).  A comparison
between perturbation theory and numerical simulations is shown in the
inset of figure~\ref{fig:y_res_param}. Again agreement is good but
fails to predict where the upper branch becomes unstable.  The
destabilization of the parametric $y-$resonance occurs in the same way
as the fundamental $y-$resonance, i.e. through parametric excitation
of the $x_-$-mode and can be analyzed following the along the lines
of the calculation in Appendix~\ref{app:destaby}.

\begin{figure}[t]
\begin{center}
\epsfig{file=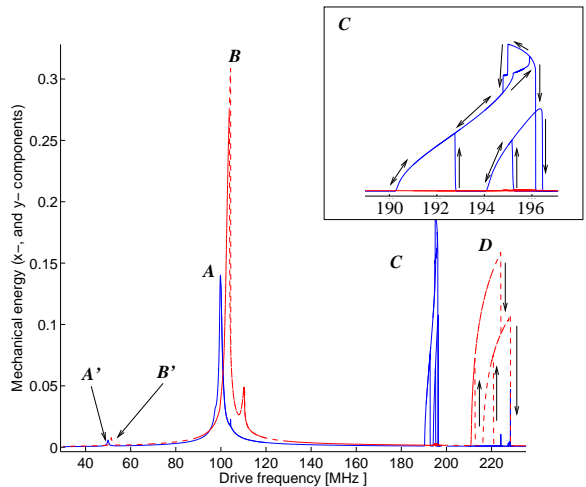,width=8cm}
\end{center}
\caption{
  (Color online) Response of a four-tube system with the same physical
  parameters as the one in figure~\ref{fig:sim_2_tubes}. The inset
  shows a close-up of the parametric transverse resonance with the
  directions of the transitions in the frequency plane indicated by
  arrows.
\label{fig:fourtub}}
\end{figure}
\begin{figure}[t]
\begin{center}
\epsfig{file=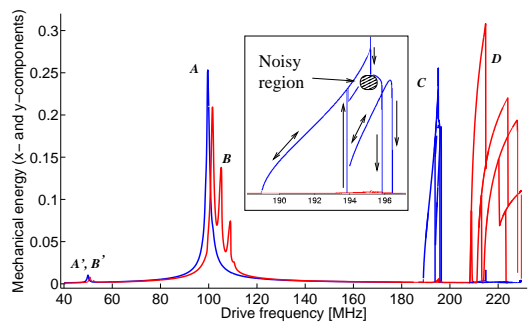,width=8cm}
\end{center}
\caption{
  (Color online) Response of an 8-tube system with the same physical
  parameters as the one in figure~\ref{fig:sim_2_tubes}. The inset
  shows a close-up of the parametric transverse resonance with the
  directions of the transitions in the frequency plane indicated by
  arrows. In the area denoted noisy region, the coupling to
  longitudinal motion causes the amplitude of transverse motion to
  oscillate in time.
\label{fig:8tubes}}
\end{figure}
\begin{figure}[t]
\begin{center}
\epsfig{file=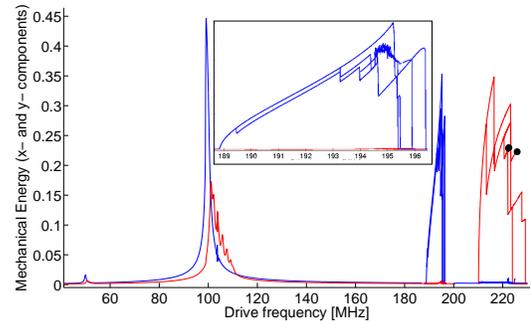,width=8cm}
\end{center}
\caption{
  (Color online) Response of a 16-tube system with the same physical
  parameters as the one in figure~\ref{fig:sim_2_tubes}. The black
  dots indicate that when the parametric longitudinal resonance became
  unstable, strong excitations of transverse modes occured that lead
  to snap-to-contact of the system.  The inset shows a close up of the
  parametric transverse resonance. In the area around 195 Mhz
  irregular behavior with high amplitude motion occurs.
\label{fig:16tub}}
\end{figure}
\section{Several oscillators \label{sec:Several}}
Having treated the two-oscillator system in some detail we now move on
to describe how the system response changes with increasing system
size. For this we use the same system parameters (geometry and bias
voltages) as those used to obtain figure~\ref{fig:sim_2_tubes} and
only change the number of tubes in the array. We have done detailed
simulations for systems with 4, 8, and 16 tubes and the corresponding
frequency responses are shown in
figures~\ref{fig:fourtub}-\ref{fig:16tub}.

The fundamental transverse resonance is not markedly affected by the
increasing array size. This is expected since here all tubes oscillate
in phase with each other.  The fundamental longitudinal resonance is
however strongly affected, the single, hysteretic peak from the
two-tube system, developing into a broad band of excited oscillation
modes. The presence of this band is reflected also in the longitudinal
parametric resonances, where the development of band structure is
present in terms of multiple branches and bifurcations in the
response. This type of behavior has been seen in parametrically driven
NEM/MEM arrays~\cite{Buks2}.  Also for the larger arrays
large amplitude excitations of longitudinal oscillations
can be destabilized due to parametric excitation of transverse
modes.  In figure~\ref{fig:16tub} two particular such points are
marked with black circles. At these points the excitation of the
transverse modes became so strong that snap-to-contact occured.

Also the parametric transverse resonance shows the development of a
band structure.  In contrast to the fundamental resonance where this
band structure is not accessible, several branches can be reached
through parametric excitation. In
figures~\ref{fig:fourtub}-\ref{fig:16tub} the insets show closeups of
the parametric transverse resonances.

While more and more modes appear as the arrays get larger, one feature
is common to all the systems.  This feature is the noisy region around
195 MHz. In this region, energy is transferred between transverse and
longitudinal modes just as in the case of the two-tube system (see
figure~\ref{fig:xparamclose}) but without destabilizing the transverse
motion.

As for the location of the resonances in the voltage-frequency plane
these do not differ appreciably from the two-tube system and the
perturbative formulas derived in the preceeding section can be used to
estimate if and where the system will be unstable to a certain
resonance.

\section{Conclusions \label{sec:conclusions}}
In order to investigate the effects of electrostatic interactions
between carbon nanotubes in NEM-resonator arrays we have studied a
simple model both analytically and numerically. We have found that,
apart from excitation (fundamental and parametric) of a band of
longitudinal modes, also parametric excitation of transverse modes is
possible.  With increasing number of resonators, these resonances
become successively more complicated and exhibit rich behavior with
several overlapping hysteresis loops, bifurcation points etc.  The
transverse modes are also responsible for destabilizing the
longitudinal modes at high amplitudes and may lead to snap to contact.
Also, the parametrically excited transverse modes, show regions of
irregular behavior coming from coupling between 
transverse and longitudinal modes.  We have shown, that the
features of the response of 1D-arrays can be understood
qualitatively through studying the simplest possible array, a two-tube
system. Also quantitative predictions based on the two tube system can
be used to obtain estimates of regions of instability towards
parametric resonances and to estimate frequency tuning.

From a technological point of view, these estimates can help in
designing array resonator systems to avoid unwanted resonances while
maintaining a high packing density. Utilizing parametric resonances
could also be a path to further increase the operation frequency in
technical applications and by tuning the bias voltages the width of
the region of instability can be tuned to an arbitrarily narrow
frequency domain. So far, only uniform arrays have been studied. For
applications, disorder must be accounted for and further studies are
needed.

\begin{acknowledgments}
  This work was supported by the Swedish Foundation for Strategic
  Research (SSF) and the EU through the Nano-RF project
  FP6-2005-028158.This publication reflects the views of the authors
  and not necessarily those of the EC. The EC is not liable for any
  use that may be made of the information contained herein.
\end{acknowledgments}

\begin{appendix}
\section{Destabilization of primary $y$-resonance\label{app:destaby}}
We here give a brief derivation of the criteria for destabilization of
the fundamental longitudinal mode through parametric excitation of the
transverse $x_-$-mode of the two-tube system. Following the same
lines, the stability of the parametric longitudinal excitation can be
analyzed.

The longitudinal vibrations are destabilized by the $x_-$-mode, which
has the equation of motion
\begin{eqnarray}
\ddot{x}_-+\gamma \dot{x}_-+\omega_0^2 x_-
=\nu\omega_0^2v^2\left[\frac{g_1^2}{(x_++x_-)^2}\right.&&\nonumber\\
-\left.\frac{g_2^2}{(x_+-x_-)^2}+2g_1g_2\frac{x_-}{(x_-^2+y^2)^{3/2}}\right].\nonumber&&
\end{eqnarray}
For small oscillations of the $x_-$-mode, the right hand side can be
approximated for large amplitudes $Y$ of the $y$-mode (recalling that
$y=y_s+Y\cos\omega_D t$) yielding the equation
$$\ddot{x}_-+\gamma \dot{x}_-+\omega_0^2 x_-=-x_-F(t)$$
where
\begin{equation}
F(t)\equiv\frac{2\nu A^2\omega_0^2v_0^2(1+2\epsilon\cos(\omega_Dt+\delta))}{(y_s+Y\cos(\omega_D t))(Y\cos(\omega_Dt)+y_s-A)^2}
\label{eq:approxyF}
\end{equation}
and we have defined $A\equiv(2+1/x_s-\alpha)^{-1}$ and $\alpha\equiv
(x_s^2+y_s^2)^{-1/2}$ respectively. Changing to action angle
variables ($x_-=X(t)\cos[\phi(t)]$, $\dot{x}_-=-X(t)\omega\sin[\phi(t)]$) and
averaging over fast variables results in the autonomous system
\begin{eqnarray}
\dot{X}&=&X\left[\frac{\left<\sin 2\phi F(t)\right>}
{2\sqrt{\omega_0^2+c_0}}-\frac{\gamma}{2}\right]\nonumber\\
\dot{\phi}&=&\sqrt{\omega_0^2+c_0}
+\frac{1}{2\sqrt{\omega_0^2+c_0}}\left<\cos2\phi F(t)\right>\nonumber,
\end{eqnarray}
where the brackets denotes the averaging $\left<\cdot\right>\equiv
(2\pi)^{-1}\int_0^{2\pi}\,\,\cdot\,d\phi$,  and we have expanded $F(t)$ in a
Fourier series $F(t)=\sum_n c_ne^{in\omega_Dt}$. At the onset of the
destabilizing $x_-$-resonance we have $\dot{\phi}=\omega_D$ and
$\phi=\omega_Dt+\vartheta$. Evaluating the averages and setting
$c_2=|c_2|e^{i\lambda}$ one finds:
\begin{eqnarray}
\dot{X}&=&X\left[|c_2|\frac{\sin(2\vartheta-\lambda)}
{2\sqrt{\omega_0^2+c_0}}-\frac{\gamma}{2}\right]\nonumber\\
\omega_D-\sqrt{\omega_0^2+c_0}&=&\frac{1}{2\sqrt{\omega_0^2+c_0}}|c_2|\cos(2\vartheta-\lambda).\nonumber
\end{eqnarray}
The region of driving frequencies where the high amplitude branch of
the $y$-mode can be destabilized by the $x_-$-mode can then be found
as $\omega_D^{\rm destab-}-<\omega_D<\omega_D^{\rm destab.+}$ with
$$\omega_D^{\rm destab.\pm}=\left[\sqrt{\omega_0^2+c_0}\pm \sqrt{\frac{|c_2|^2}{4(\omega_0^2+c_0)}-\frac{\gamma^2}{4}}\right].$$
Using the expression (\ref{eq:approxyF}) the Fourier coefficients $c_0$ and $c_2$ can
be evaluated exactly: 
\begin{widetext}
\begin{eqnarray}
\frac{c_0}{\omega_0^2}&=&2\nu\left[\frac{Y+2(A-y_s)\epsilon\cos\delta}{Y\sqrt{-Y^2+(A-y_s )^2}}-\frac{Y+2\epsilon y_s\cos\delta}{Y\sqrt{-Y^2+y_s ^2}}+\frac{A(A-y_s )+2AY\epsilon\cos\delta}{\left(-Y^2+(A-y_s )^2\right)^{3/2}}\right],
\end{eqnarray}
\begin{eqnarray}
&&\frac{c_2}{\omega_0^2}=2\nu\left[\frac{\left(Y^2+(A-y_s )y_s +A(2Y)\epsilon\cos\delta\right)}{\left(-Y^2+(A-y_s )^2\right)^{3/2}}-2\frac{Y(A-y_s)y_s-(y_s Y^2-2(A+y_s)(A-y_s)^2)\epsilon\cos\delta}{Y^3 \sqrt{-Y^2+(A-y_s )^2}}\right]\nonumber\\
&&+\left.\frac{ \left(Y^2-2 y_s ^2)(Y-2y_s\epsilon\cos\delta)\right)}{Y^3\sqrt{-Y^2+y_s ^2}}+\frac{4A^2\epsilon\cos\delta}{Y^3}\right].
\end{eqnarray}
\end{widetext}
After solving the frequency response equation for the $y$-resonance,
$Y$ and $\cos\delta$ can be found and the expressions can be
evaluated, thus determining whether or not parametric excitation of
the $x_-$-mode will occur.

\end{appendix}

\end{document}